\documentstyle[epsf,aps,twocolumn]{revtex}
\begin{document}
\title{\hspace{5cm} {\em Solid State Communications} {\bf 106}
{\em (1998) 451--455}\\
\vskip 2cm
ON ELECTROSTRICTION OF A GRANULAR SUPERCONDUCTOR}
\author{Sergei A. Sergeenkov}
\address{
Bogoliubov Laboratory of
Theoretical Physics, Joint Institute for Nuclear Research,\\
141980 Dubna, Moscow Region, Russia}
\vskip 2.5cm
\date{\today}
\address{~}
\address{
\centering{
\begin{minipage}{16cm}
Zero-temperature field-induced polarization, supercurrent density,
and the related electrostriction (ES) of a granular superconductor are
calculated
within a model of 3D Josephson junction arrays. Both the "bulk-modulus-
driven ES" (the change of the sample's volume in the free energy upon
the applied stress) and the "change-of-phase ES" (due to the stress
dependence of the weak-links-induced polarization) are considered.
In contrast to magnetostriction of a granular superconductor,
its electroelastic behavior is predicted to be dominated by
the former contribution for all applied fields.
\end{minipage}
}}
\maketitle

\narrowtext

Some attention was given recently to rather peculiar electric-field
induced phenomena, either observed experimentally (like a substantial
critical current enhancement~\cite{ref1,ref2,ref3}) or predicted to
occur (like a possibility of magnetoelectric effect due to the
Dzyaloshinski-Moria type coupling between an applied electric field
and an effective magnetic field of circulating Josephson
currents~\cite{ref4}) in granular superconductors and attributed to
their weak-link structure.
At the same time, as compared to the magnetoelastic behavior of
superconducting materials (dominated either by a vortex
response~\cite{ref5,ref6,ref7,ref8} or by weak-links
structure~\cite{ref9}), their electroelastic behavior still remains
to be properly addressed.

In the present communication, another interesting pnenomenon related
to the modification of the sample's weak-links structure in an applied
electric field is discussed. Namely, we consider a possible role of
Josephson junctions in low-temperature behavior of the field-induced
polarization and the related electroelastic properties of granular
superconductors.

As is well-known~\cite{ref10}, the change of the free energy of a
superconductor in the presence of an external electric field $E$ reads
\begin{equation}
\Delta F(E_i)\equiv F(0)-F(E_i)=V\int_{0}^{E_i}dEP(E),
\end{equation}
where $P(E)$ is the electric polarization of a granular superconductor
at zero temperature (see below), $V$ its volume, and the internal field
$E_i$ is related to the applied field $E$ via an effective dielectric
constant $\epsilon$, namely~\cite{ref10} $E_i=E/\epsilon$.
When a superconductor is under the influence of an external
(homogeneous) stress $\sigma$, the above free energy results in the
associated strain component (in what follows, we consider only a strain
component $U$ normal to the applied electric field $E$)
\begin{equation}
U=\frac{1}{V}\left (\frac{\partial \Delta F}{\partial \sigma}\right ).
\end{equation}
Neglecting a possible change of the effective dielectric constant
$\epsilon$ with the stress, Eqs.(1) and (2) give rise to the following
two main contributions to the electrostrictive (ES) strains, namely

(a) the "bulk-modulus-driven ES" due to the change in the free energy
arising from the stress dependence of the sample volume
\begin{equation}
U_{BMD}\equiv \left (\frac{1}{V}\frac{\partial \Delta F}{\partial
\sigma}\right )_P=\left (\frac{1}{V}\frac{\partial V}{\partial \sigma}
\right )\int_{0}^{E_i}dEP(E);
\end{equation}

(b) the "change-of-phase ES" due to the stress dependence of the
polarization via the Josephson junction effective surface (see below)
\begin{equation}
U_{PH}\equiv \left (\frac{1}{V}\frac{\partial \Delta F}{\partial
\sigma}\right )_V=\int_{0}^{E_i}dE\frac{\partial P(E)}{\partial \sigma}.
\end{equation}
To proceed, we need an explicit form of the induced polarization $P(E)$.
And to this end, we employ the model of a granular superconductor based
on the well-known tunneling Hamiltonian (see, e.g., Ref.~\cite{ref11})
\begin{equation}
{\cal H}=\sum_{ij}^NJ_{ij}[1-\cos \phi _{ij}(t)],
\end{equation}
where
\begin{equation}
\phi _{ij}(t)=\phi _{ij}(0)+\omega _{ij}(\vec E)t,
\end{equation}
with
\begin{equation}
\omega _{ij}(\vec E)=\frac{2e}{\hbar }\vec E \vec r_{ij},
\end{equation}
and
\begin{equation}
\phi _{ij}(0)=\phi _i-\phi _j, \qquad \vec r_{ij}=\vec r_i-\vec r_j,
\end{equation}
which describes an interaction between superconducting grains (with
phases $\phi _i(t)$), arranged in a random three-dimensional (3D)
lattice with coordinates $\vec r_i=(x_i,y_i,z_i)$. The grains are
separated by insulating boundaries producing Josephson coupling with
energy $J_{ij}=J$. The system is under the influence of an external
electric field $\vec E=(E,0,0)$.

The corresponding pair polarization operator within the model
reads~\cite{ref12}
\begin{equation}
\vec p=\sum_{i}^Nq_i \vec r_i,
\end{equation}
where $q_i =-2en_i$ with $n_i$ the pair number operator, and $r_i$ is
the coordinate of the center of the grain.

In view of Eqs.(5)-(9), and taking into account a usual "phase-number"
commutation relation, $[\phi _i,n_j]=i\delta _{ij}$, the evolution of
the polarization operator obeys the equation of motion
\begin{equation}
\frac{d\vec p}{dt}=\frac{1}{i\hbar}\left[ \vec p,{\cal H}\right ]=
\frac{2e}{\hbar}\sum_{ij}^NJ\sin \phi _{ij}(t)\vec r_{ij}
\end{equation}
Resolving the above equation, we arrive at the following mean value of
the field-induced polarization
\begin{eqnarray}
\vec P(\vec E)&\equiv &\frac{1}{V}\overline {<\vec p(t)>} \nonumber \\
&=&\frac{2eJ}
{\hbar \tau V} \int\limits_{0}^ {\tau }dt \int
\limits_{0}^{t}dt'\sum_{ij}^N <\sin \phi _{ij}(t')\vec r_{ij}>,
\end{eqnarray}
where $<...>$ denotes a configurational averaging over the grain
positions, while the bar means a temporal averaging with a
characteristic time $\tau$ (see below).

To limit ourselves with field-induced polarization effects only,
we assume that in a zero electric field, $\vec P\equiv 0$, and thus
$\phi _{ij}(0)\equiv0$.

To obtain an explicit form of the field dependence of polarization,
let us consider a site-type positional disorder allowing for weak
displacements of the grain sites from their positions of the original
$3D$ lattice, i.e., within a radius $d\approx \sqrt{S}$ ($S$ is an
effective surface of grain-boundary Josephson junction) the new
position is chosen randomly according to the normalized (separable)
distribution function $f(\vec r)=f(x)f(y)f(z)$. It can be shown~\cite{ref13}
that
the main {\it qualitative} results of this paper do not depend on the
particular choice of the probability distribution function. Hence,
assuming, for simplicity, a normalized exponential distribution law,
$f(x)=(1/d)e^{-x/d}$ (where $x>0$ and $\int_0^{\infty}dxf(x)=1$), we
find that
the electric field $\vec E=(E,0,0)$ (applied along the $x$-axis)
will produce a non-vanishing longitudinal (along $x$-axis) polarization
vector $\vec P=(P,0,0)$ with
\begin{equation}
P(E)=P_0G(E/E_0),
\end{equation}
where
\begin{equation}
G(z)=\frac{1}{z}\left (1-\frac{\arctan z}{z}\right )
\end{equation}
Here $P_0=2JN/E_0V$, $E_0=\hbar /ed\tau$, and $z=E/E_0$.

At the same time, as is well-known~\cite{ref12}, the supercurrent density
through the Josephson junction between grains $i$ and $j$ is related to
the polarization operator $\vec p$ as follows (see Eq.(10))
\begin{equation}
\vec j_s(\vec E)\equiv \frac{1}{V}\overline{\left <\frac{d\vec p}{dt}\right >}
\end{equation}
Repeating the above-discussed averaging procedure, we find for
the change of the longitudinal part of the supercurrent density in applied
electric field
\begin{equation}
j_s(E)=j_0D(E/E_0),
\end{equation}
where
\begin{equation}
D(z)=\frac{z}{1+z^2}
\end{equation}
with $j_0=2eJNd/\hbar V$, and $z=E/E_0$.

Figures 1 and 2 show the field-induced behavior of the normalized
polarization $P(E)/P_0$ and supercurrent density $j_s(E)/j_0$, calculated
according to Eqs.(12) and (15), respectively. Notice a rather pronounced
peak at $E/E_0\simeq 2$ for both dependences. Assuming $d\approx 1\mu m$
and $\tau \approx 10^{-16}s$ for an average grain size and a low-temperature
estimate of the Josephson tunneling time of Cooper pairs through an
insulating barrier in applied electric field~\cite{ref14} in $YBCO$,
we get $E_0\simeq 10^7 V/m$ for the estimate of the model
characteristic field, which is very close to the typical applied field
values where the critical current of ceramic samples was
found~\cite{ref1,ref2,ref3} to reach its maximum. Besides, assuming
$V\simeq Nd^3$ and taking into account that the Josephson energy in $YBCO$
is $J/k_B\simeq 90K$, we obtain quite a reasonable estimate for the
model characteristic critical current density $j_0$, typical for ceramic
samples~\cite{ref1,ref2}. Namely $j_0\simeq 2eJ/\hbar d^2\simeq 10^6A/m^2$.
Let us briefly comment on the temporal averaging (used in Eq.(11)) and
discuss the relationship between the characteristic time
$\tau =\hbar /eE_0d$ and period of oscillations $T(E)$. The latter
is defined via Eq.(7) as $T(E)=2\pi /<\omega _{ij}(E)>$, where
$<\omega _{ij}(E)>=2eEd/\hbar$. Thus, depending on the strength of an
applied electric field, the period of oscillations $T(E)$ can be larger or
smaller than the tunneling time $\tau$. In particular, high-field region
$E\geq E_0$ where the most interesting effects take place, is
characterized by faster oscillations with the period $T(E) \leq \tau$,
as compared with low-field behavior. In addition, we can compare $\tau$ with
a zero-field (and low-temperature) Josephson tunneling time
$\tau _0\simeq \hbar /J$ which in $YBCO$ gives $\tau _0\simeq 10^{-13}s$.
Hence, at low applied fields (when $E\ll E_0$) $T(E)\simeq \tau _0$ while in
high-field regime (when $E\gg E_0$) $T(E)\ll \tau \ll \tau _0$.
Furthermore, according to Eqs.(3) and (12), the explicit form of the
"bulk-modulus-driven ES" contribution is as follows
\begin{equation}
U_{BMD}(E_i)=U_0G_{BMD}(E_i/E_0),
\end{equation}
where
\begin{equation}
G_{BMD}(z)=zG(z)+\ln {\sqrt{1+z^2}}
\end{equation}
Here $U_0=\kappa E_0P_0$ with $\kappa =-\partial \ln V/\partial
\sigma$ being the compressibility coefficient.

To evaluate the "change-of-phase ES" contribution $U_{PH}(E_i)$, we
have to account for the stress dependence of the effective surface
$S(\sigma )$ of grain-boundary Josephson junctions which
was found~\cite{ref15} experimentally to decrease with the applied
stress. Using the following chain of evident relations,
\begin{eqnarray}
\frac{\partial P}{\partial \sigma}&=&\frac{\partial P_0}{\partial
\sigma}G+P_0\frac{\partial G}{\partial \sigma},\nonumber\\
\frac{\partial P_0}{\partial \sigma}&=&\frac{\partial P_0}{\partial V}
\frac{\partial V}{\partial \sigma}=P_0\kappa ,\nonumber\\
\frac{\partial G}{\partial \sigma}&=&\frac{\partial G}{\partial E_0}
\frac{\partial E_0}{\partial \sigma}, \nonumber\\
\frac{\partial E_0}{\partial \sigma}&=&\frac{\partial E_0}{\partial S}
\frac{\partial S}{\partial \sigma},
\end{eqnarray}
and assuming that the sample volume $V$ and the projected area $S$ are
related in a usual way, $S\approx V^{2/3}$, we obtain from Eqs.(4),
(12) and (19) for the "change-of-phase ES"
\begin{equation}
U_{PH}(E_i)=-\frac{1}{3}U_0G_{PH}(E_i/E_0),
\end{equation}
where
\begin{equation}
G_{PH}(z)=zG(z)+2G_{BMD}(z)
\end{equation}
Figure 3 summarizes the predicted behavior of the two considered
contributions (calculated according to Eqs.(17) and (20)), along with
their total effect (solid line) on induced electrostriction. Notice
that in contrast to the magnetostrictive behavior of a granular
superconductor (considered in Ref.~\cite{ref9}), its electroelastic
properties are completely dominated by the "bulk-modulus-driven"
contribution over the whole region of applied fields. It would be
interesting to verify the above-predicted behavior experimentally.

In conclusion, a low-temperature field-induced electroelastic behavior
of a granular superconductor was considered within a model of 3D
Josephson junction arrays. The "bulk-modulus-driven" contribution to
the electrostriction (ES) of a granular superconductor (related to the
change of the sample's volume in the free energy upon the applied
stress) was shown to dominate over the "change-of-phase" ES (related
to the stress dependence of the weak-links-induced polarization)
for all applied fields.

\acknowledgments
This work was financially supported by the Brazilian agency CNPq.

\newpage
\begin{figure}
\epsfxsize=8cm
\centerline{\epsffile{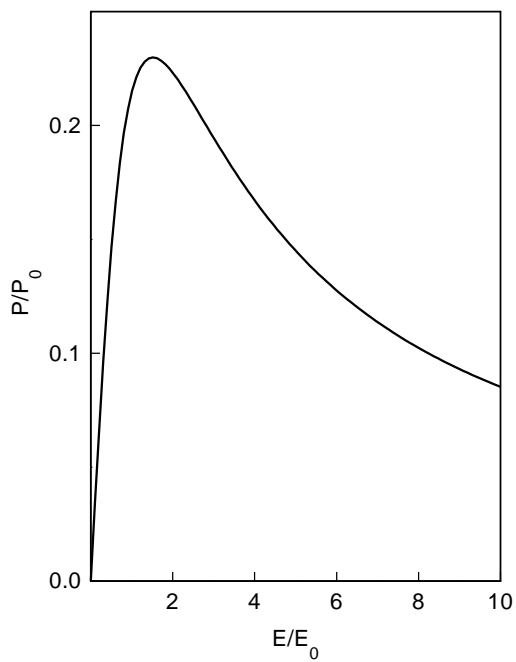}   }
\caption{The behavior of the induced polarization $P/P_0$ in applied
electric field $E/E_0$.}
\end{figure}

\begin{figure}
\epsfxsize=8cm
\centerline{\epsffile{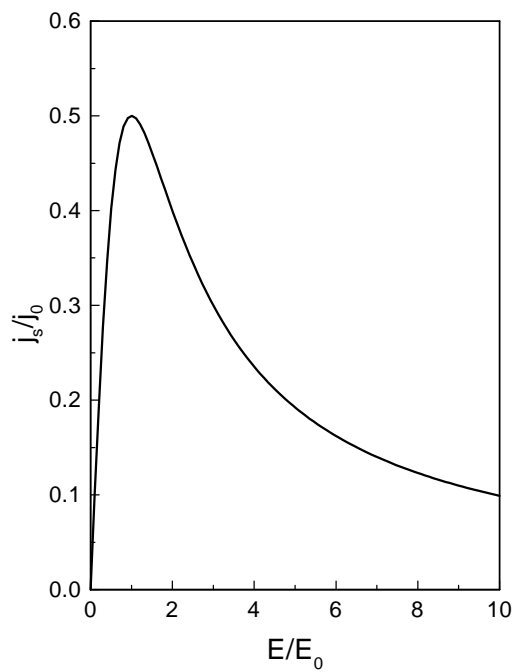}   }
\caption{The behavior of the induced supercurrent density $j_s/j_0$ in
applied electric field $E/E_0$.}
\end{figure}

\begin{figure}
\epsfxsize=8cm
\centerline{\epsffile{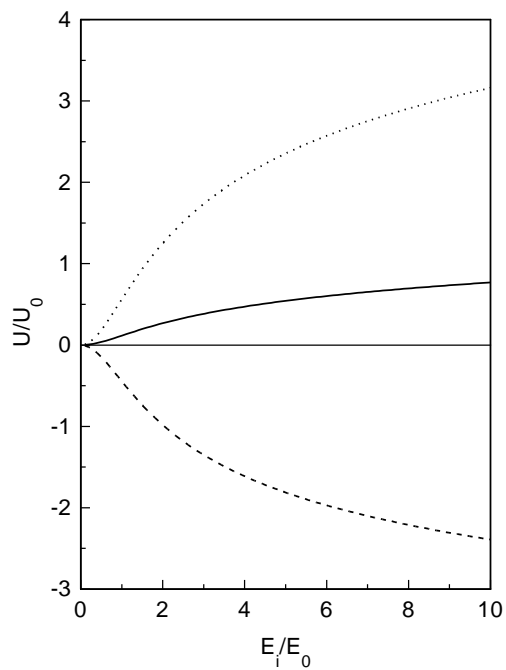}   }
\caption{The "bulk-modulus-driven" $U_{BMD}/U_0$
(dotted line), "change-of-phase" $U_{PH}/U_0$ (dashed line),
and the total $(U_{PH}+U_{BMD})/U_0$ (solid line) contributions
to the induced electrostriction
vs internal electric field $E_i/E_0$.}
\end{figure}


\begin{thebibliography}{99}
\bibitem{ref1} Smirnov, B.I., Orlova, T.S. and Krishtopov, S.V.,
{\it Phys. Solid State}, {\bf 35}, 1993, 1118.
\bibitem{ref2} Orlova, T.S. and Smirnov, B.I., {\it Supercond. Sci.
Techn.}, {\bf 7}, 1994, 899.
\bibitem{ref3} Rakhmanov, A.L. and Rozhkov, A.V., {\it Physica C},
{\bf 267}, 1996, 233.
\bibitem{ref4} Sergeenkov, S. and Jos\'e, J., {\it Bull. Am. Phys.
Soc.}, {\bf 42}, 1997, 121.
\bibitem{ref5} Brandli, G., {\it Phys. Kondens. Mater.}, {\bf 11},
1970, 93.
\bibitem{ref6} Braden, M., Bohm, P., Seidler, F., Kalenborn, H. and
Wohlleben, D., {\it Z. Phys.}, {\bf B79}, 1990, 173.
\bibitem{ref7} del Moral, A., Ibarra, M.R., Algarabel, P.A. and
Arnaudy, J.I., {\it Physica C}, {\bf 161}, 1989, 48.
\bibitem{ref8} Oliveira, N.F., Nicholls, J.T., Shapira, Y., Dresselhaus, G.,
Dresselhaus, M.S., Picone, P.J., Gabbe, D.R. and Jenssen, H.P.,
{\it Phys. Rev.}, {\bf B39}, 1989, 2898.
\bibitem{ref9} Sergeenkov, S. and Ausloos, M., {\it Phys. Rev.},
{\bf B48}, 1993, 604.
\bibitem{ref10}  Landau, L.D. and Lifshitz, E.M., {\em Electrodynamics
of Continuous Media} (Pergamon Press, Oxford, 1960).
\bibitem{ref11} Ebner, C. and Stroud, D., {\it Phys. Rev.},
{\bf B31}, 1985, 165.
\bibitem{ref12} Lebeau, C., Rosenblatt, J., Robotou, A. and Peyral, P.,
{\it Europhys. Lett.}, {\bf 1}, 1986, 313.
\bibitem{ref13} Sergeenkov, S., {\it J. Phys. I France}, {\bf 7},
1997, 1175.
\bibitem{ref14} Mannhart, J., {\it Supercond. Sci. Techn.}, {\bf 9},
1996, 49.
\bibitem{ref15} Stankowski, J., Czyzak, B., Krupski, M., Baszynski, J.,
Datta, T., Almasan, C. and Shang, Z., {\it Physica C}, {\bf 160},
1989, 170.
\end{thebibliography}
\end{document}